\documentclass[bstyle]{JWPublic}

\usepackage{array}
\usepackage{amssymb,latexsym,amsfonts,amsmath}
\usepackage{graphicx}
\usepackage{mathrsfs}
\usepackage{natbib}
\bibpunct{(}{)}{,}{a}{}{,}

\begin{document}

%
%
%
\mainmatter

\ContribChap{Bayesian Inference}
	       {Christian P. Robert${}^{1,3}$, Jean-Michel Marin${}^{2,3}$ and Judith Rousseau${}^{1,3}$}
	       {${}^1$Universit\'e Paris-Dauphine, ${}^2$Universit\'e de Montpellier 2, and ${}^3$CREST, INSEE, Paris}

\section{Introduction}\label{sub:intro}

This chapter provides a overview of Bayesian inference, 
mostly emphasising that it is a universal {\em method for summarising uncertainty and making
estimates and predictions using probability statements conditional on observed data and an assumed model}
\citep{gelman:2008}. The Bayesian perspective is thus applicable to all aspects of statistical inference, 
while being open to the incorporation of information items resulting from earlier experiments and from 
expert opinions.

We provide here the basic elements of Bayesian analysis 
when considered for standard models, refering to \cite{marin:robert:2007} and
to \cite{robert:2007} for book-length entries.\footnote{The chapter borrows
heavily from Chapter 2 of \cite{marin:robert:2007}.} In the following, we refrain from embarking upon philosophical discussions
about the nature of knowledge \citep[see, e.g.,][Chapter 10]{robert:2007}, opting instead
for a mathematically sound presentation of an eminently practical statistical methodology. 
We indeed believe that the most convincing arguments for adopting
a Bayesian version of data analyses are in the versatility of this 
tool and in the large range of existing applications, rather than in
those polemical arguments (for such perspectives, see, e.g., \citealp{jaynes:2003} and
\citealp{mackay:2002}).

\section{The Bayesian argument}\label{sec:tooBayes}

\subsection{Bases}\label{sub:baz}

We start this section with some notations about the statistical model that may appear to be
over-mathematical but are nonetheless essential.

Given an independent and identically distributed (iid) sample $\mathscr{D}_n=(x_1,\ldots,x_n)$
from a density $f_\theta$, with an unknown parameter $\theta\in\Theta$, like the mean
$\mu$ of the benchmark normal distribution, the associated {\em likelihood function}
is\index{Independent, identically distributed (iid)}\index{Likelihood}
$$
\ell(\theta|\mathscr{D}_n) = \prod_{i=1}^n f_\theta(x_i)\,.
$$
This quantity is a fundamental entity for the analysis of the information
provided about the parameter $\theta$ by the sample $\mathscr{D}_n$, and Bayesian analysis
relies on this function to draw inference on $\theta$.\footnote{Resorting to an abuse of notations, we
will also call $\ell(\theta|\mathscr{D}_n)$ our {\em statistical model}, even though the distribution
with the density $f_\theta$ is, strictly speaking, the true statistical model.} Since all models are approximations 
of reality, the choice of a sampling model is wide-open for criticisms (see, e.g., \citealp{templeton:2008}), but those
criticism go far beyond Bayesian modelling and question the relevance of completely built models for drawing inference 
or running predictions. We will therefore address the issue of model assessment later in the chapter.

The major input of the Bayesian perspective, when compared with a standard likelihood approach, is
that it modifies the likelihood---which is a simple function of $\theta$---into a {\em posterior} distribution
on the parameter $\theta$---which is a probability
distribution on $\Theta$ defined by\index{Posterior}\index{Bayesian!posterior|see{Posterior}}
\begin{equation}\label{eq:poste}
\pi(\theta|\mathscr{D}_n) = \frac{\ell(\theta|\mathscr{D}_n)\pi(\theta)}{
\int\,\ell(\theta|\mathscr{D}_n)\pi(\theta)\,\hbox{d}\theta}\,.
\end{equation}
The factor $\pi(\theta)$ in \eqref{eq:poste} is called the {\em prior} (often omitting the qualificative {\em density}) and
it necesssarily has to be determined to start the analysis. A primary motivation for introducing
this extra-factor is that the prior distribution summarizes the {\em prior information}
on $\theta$; that is, the knowledge that is available on $\theta$ {\em prior} to the
observation of the sample $\mathscr{D}_n$. However, the choice of $\pi(\theta)$
is often decided on practical or computational grounds rather than on strong
subjective beliefs or on overwhelming prior information. As will be discussed later,
there also exist less subjective\index{Prior!subjective}\index{Bayesian!prior|see{Prior}}
choices, made of families of so-called {\em noninformative priors}.

The radical idea behind Bayesian modelling is thus that the {\em uncertainty} on the unknown parameter $\theta$ is
more efficiently modelled as {\em randomness} and consequently that the probability distribution $\pi$ is needed on
$\Theta$ as a reference measure.  In particular, the distribution $P_\theta$ 
of the sample $\mathscr{D}_n$ then takes the meaning of a probability distribution on $\mathscr{D}_n$ 
that is conditional on [the event that the parameter takes] the value $\theta$, i.e.~$f_\theta$ is the 
{\em conditional} density of $x$ given $\theta$. The above likelihood offers the dual interpretation of the
probability density of $\mathscr{D}_n$ {\em conditional} on the parameter $\theta$, with the additional indication
that the observations in $\mathscr{D}_n$ are independent given $\theta$. The numerator of \eqref{eq:poste} is therefore
the joint density on the pair $(\mathscr{D}_n,\theta)$ and the (standard probability calculus) Bayes theorem provides
the {\em conditional} (or {\em posterior\/}) distribution of the parameter $\theta$ given the sample $\mathscr{D}_n$ as \eqref{eq:poste},
the denominator being called the marginal (likelihood) $m(\mathscr{D}_n)$.

There are many arguments which make such an approach compelling. When defining a probability
measure on the parameter space $\Theta$, the Bayesian approach endows notions such as {\em the
probability that $\theta$ belongs to a specific region} with a proper meaning 
and those are particularly relevant when designing measures of uncertainty like 
confidence regions or when testing hypotheses. Furthermore, the posterior distribution
\eqref{eq:poste} can be interpreted as the actualisation of the knowledge (uncertainty) on the parameter after observing
the data. At this early stage, we stress that the Bayesian perspective does not state that the model within
which it operates is the ``truth", no more that it believes that the corresponding prior distribution $\pi$
it requires has a connection
with the ``true" production of parameters (since there may even be no parameter at all). It simply provides an inferential machine that
has strong optimality properties under the right model and that can similarly be evaluated under any other well-defined alternative model.
Furthermore, the Bayesian approach includes techniques to check prior beliefs as well as statistical models \citep{gelman:2008},
so there seems to be little reason for not using a given model at an
earlier stage even when dismissing it as ``un-true" later (always in favour of another model).

\subsection{Bayesian analysis in action}\label{sub:BayAct}
The operating concept that is at the core of Bayesian analysis is that one should provide an
inferential assessment {\em conditional on the realized value of $\mathscr{D}_n$},
and Bayesian analysis gives a proper probabilistic meaning to this conditioning
by allocating to $\theta$ a (reference) probability (prior) distribution $\pi$.
Once the prior distribution is selected, Bayesian inference formally is
``over"; that is, it is completely determined since the estimation, testing,
prediction, evaluation, and any other inferential procedures are automatically provided by the prior
and the associated loss (or penalty) function.\footnote{Hence the concept, introduced above, of
a complete inferential machine.} For instance, if estimations $\hat\theta$ 
of $\theta$ are evaluated via the quadratic loss function\index{Loss function!quadratic}
$$
\mbox{L} (\theta,\hat\theta) = \| \theta-\hat\theta \|^2,
$$
the corresponding Bayes procedure is the {\em expected} value of $\theta$
under the posterior distribution,\index{Bayesian!decision procedure}
\begin{equation}\label{eq:pomean}
\hat\theta = \int \theta \,\pi(\theta|\mathscr{D}_n)\,\hbox{d}\theta
            = \frac{\int \theta\,\ell(\theta|\mathscr{D}_n)\,\pi(\theta)\,\hbox{d}\theta}
                   {m(\mathscr{D}_n)}\,,
\end{equation}
for a given sample $\mathscr{D}_n$. For instance, observing a frequency $38/58$ of survivals
among $58$ breast-cancer patients and assuming a binomial $\mathcal{B}(58,\theta)$ with a
uniform $\mathcal{U}(0,1)$ prior on $\theta$ leads to the Bayes estimate
$$
\widehat{\theta}= \dfrac{\int_0^1 \theta {58 \choose 38} \theta^{38} (1-\theta)^{20}\,\text{d}\theta}
                       {\int_0^1        {58 \choose 38} \theta^{38} (1-\theta)^{20}\,\text{d}\theta}
	       = \dfrac{38+1}{58+2}\,,
$$
since the posterior distribution is then a beta ${\mathcal B}e(38+1,20+1)$ distribution.

When no specific loss function is available, the estimator \eqref{eq:pomean} is often used
as a default estimator, although alternatives also are available. For 
instance,\index{Maximum a posteriori}\index{MAP|see{Maximum a posteriori}}
the {\em maximum a posteriori estimator} (MAP) is defined as
\begin{equation}\label{eq:MAP}
  \hat\theta = \arg\max_\theta \pi(\theta|\mathscr{D}_n)
                          = \arg\max_\theta \pi(\theta)\ell(\theta|\mathscr{D}_n),
\end{equation}
where the function to maximize is usually provided in closed form. However, numerical
problems often make the optimization involved in finding the MAP far from trivial.
Note also here the similarity of (\ref{eq:MAP}) with the maximum
likelihood estimator (MLE): The influence of the prior
distribution $\pi(\theta)$ progressively disappears with the number
of observations, and the MAP estimator recovers the asymptotic properties
of the MLE. See \cite{schervish:1995} 
for more details on the asymptotics of Bayesian estimators. 

\begin{figure}
\begin{minipage}[b]{0.4\linewidth}
\begin{verbatim}
                surviv
    age  malign yes no
under 50     no  77 10
            yes  51 13
   50-69     no  51 11
            yes  38 20
above 70     no   7  3
            yes   6  3
\end{verbatim}
\vspace{.8cm}
\end{minipage}
\hspace{0.5cm}
\begin{minipage}[b]{0.6\linewidth}
\includegraphics[width=.8\textwidth]{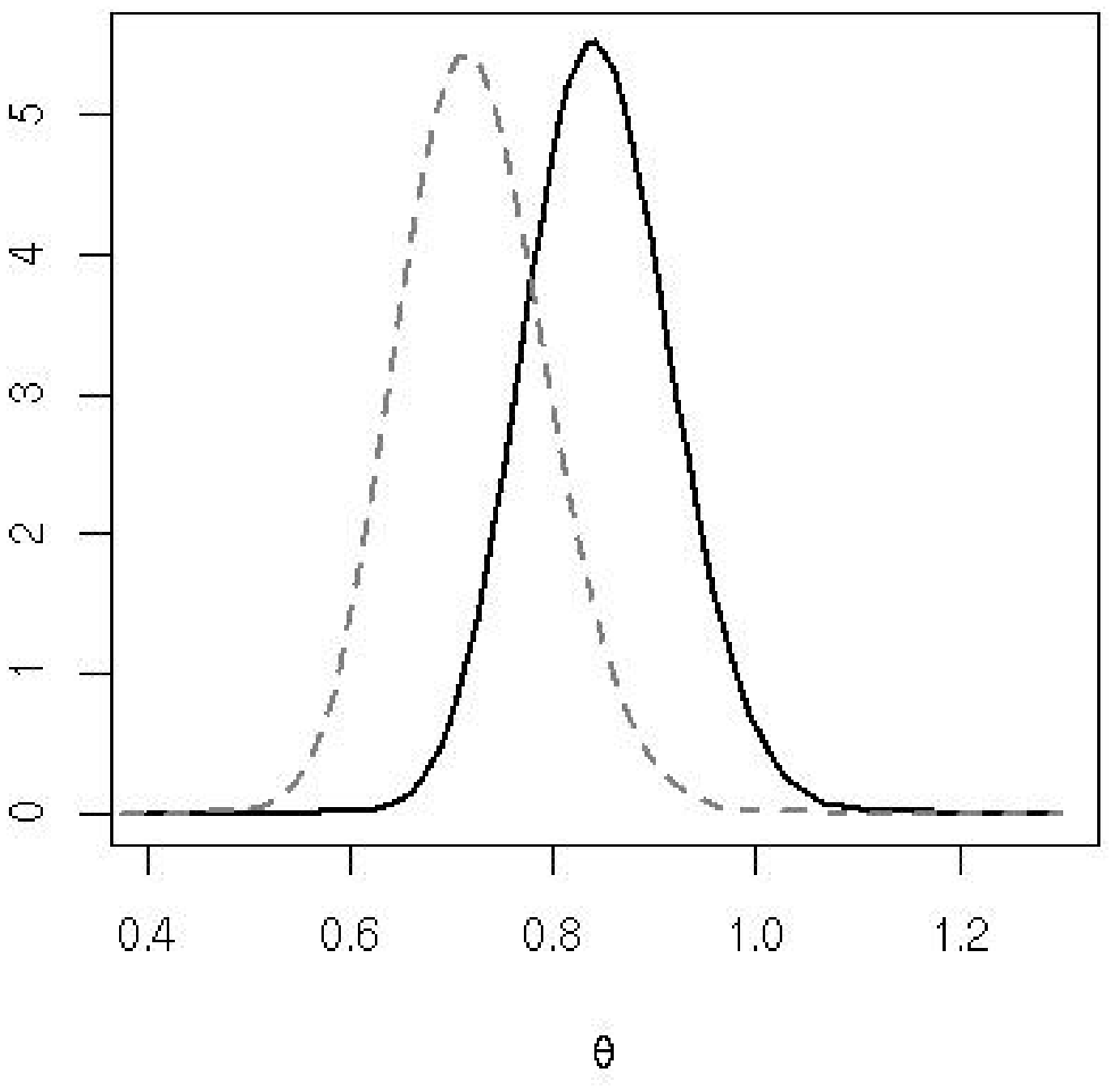}
\end{minipage}
\caption{\label{fig:cancer}
{\em (left)} Data describing the survival rates of some breast-cancer patients \citep{bishop:fienberg:holland:1975} and
{\em (right)} representation of two gamma posterior distributions differentiating between malignant {\em (dashes)}
versus non-malignant {\em (full)} breast cancer survival rates.}
\end{figure}

As an academic example, consider the contingency table provided in Figure \ref{fig:cancer} on survival rate for breast-cancer patients
with or without malignant tumours, extracted from \cite{bishop:fienberg:holland:1975}, the 
goal being to distinguish between the two types of tumour in terms of survival probability. 
We then consider each entry of the table on the number of survivors (first column of figures in 
Figure \ref{fig:cancer} to be independently Poisson distributed $\mathcal{P}(N_{it}\theta_i)$, 
where $t=1,2,3$, denotes the age group, $i=1,2$, the tumor group, distinguishing between malignant $(i=1)$ and non-malignant $(i=2)$,
      and
      $N_{it}$ is the total number of patients in this age group and for this type of tumor.
Therefore, denoting by $x_{it}$ the number of survivors in age group $t$ and tumor group $i$, the corresponding density is
$$
f_{\theta_i}(x_{it}|N_{it}) = e^{-\theta_i N_{it}}\frac{ (\theta N_{it})^{x_{it}}}{x_{it}!}, \quad x\in \mathbb{N}\,.
$$
The corresponding likelihood on $\theta_i$ $(i=1,2)$ is thus
$$
L(\theta_i|\mathcal{D}_3) = \prod_{t=1}^3 (\theta_i N_{ti})^{x_{ti}} \exp\{-\theta_i N_{ti}\}
$$
which, under an $\theta\sim\mathcal{E}xp(2)$ prior, leads to the posterior
$$
\pi(\theta_i|\mathcal{D}_3)\propto \theta^{x_{1i}+x_{2i}+x_{3i}} \exp\left\{-\theta_i(2+N_{1i}+N_{2i}+N_{3i})\right\}
$$
i.e.~a Gamma $\Gamma(x_{1i}+x_{2i}+x_{3i}+1,2+N_{1i}+N_{2i}+N_{3i})$ distribution. The choice of the (prior) exponential parameter 
corresponds to a prior estimate of $50\%$ survival probability over the period.\footnote{Once again, this is an academic example. 
The prior survival probability would need to be assessed by a physician in a real life situation.} In the case of the non-malignant breast cancers, 
the parameters of the (posterior) Gamma distribution are $a=136$ and $b=161$, while, for the malignant cancers, 
they are $a=96$ and $b=133$. Figure \ref{fig:cancer} shows the difference between both posteriors, the non-malignant 
case being stochastically closer to $1$, hence indicating a higher survival rate. 
(Note that the posterior in this figure gives some weight to values of $\theta$
larger than $1$. This drawback can easily be fixed by truncated the exponential prior at $1$.)

\subsection{Prior distributions}\label{sub:prior}

The selection of the prior distribution is an important issue in
Bayesian modelling. When prior information is available about the
data or the model, it can be used in building the prior, 
and we will see some illustrations of this recommendation
in the following chapters. In many situations,
however, the selection of the prior distribution is quite delicate in the absence
of reliable prior information, and generic solutions must be chosen instead.
Since the choice of the prior distribution has a considerable influence on
the resulting inference, this choice must be conducted with the utmost care.
It is indeed straightforward to come up with examples where a particular choice of the
prior leads to absurd decisions. Hence, for a Bayesian analysis to be sound the prior distribution needs to be well-justified.
Before entering into a brief description of some existing approaches of constructing prior distributions, note that, as part of model
checking, every Bayesian analysis needs to assess the influence of the choice of the prior, for instance through
a sensitivity analysis.  Since the prior distribution models the knowledge (or uncertainty) prior to the observation of
the data, the sparser the prior information is, the flatter the prior should be.
There actually exists a category of priors whose primary aim is to minimize the
impact of the prior selection on the inference: They are called {\em noninformative}
priors and we will detail them below.

When the sample model is from\index{Prior!selection}\index{Exponential family}
an exponential family of distributions\footnote{This covers most of the standard statistical distributions,
see \cite{lehmann:casella:1998} or \cite{robert:2007}.} with densities of the form
$$
f_\theta(x) = h(x)\,\exp \left\{ \theta\cdot R(x) - \Psi(\theta) \right\},
\qquad \theta,R(x) \in\mathbb{R}^p\,,
$$
where $\theta\cdot R(x)$ denotes the canonical scalar product in $\mathbb{R}^p$, there 
exists an associated class of priors called the {\em class of conjugate priors}, of the
form
$$
\pi(\theta|\xi,\lambda) \propto \exp \left\{ \theta\cdot \xi - \lambda \Psi(\theta) \right\}\,,
$$
which are parameterized by two quantities, $\lambda>0$ and $\xi$, $\lambda\xi$ being of the same nature as $R(y)$.
These parameterized prior distributions on $\theta$ are appealing for the simple computational reason that
the posterior distributions are exactly of the same form as the prior distributions; that is, they can be written as
\begin{equation}\label{eq:conpri}
 \pi(\theta|\xi^\prime(\mathcal{D}_n),\lambda^\prime(\mathcal{D}_n))\,,
\end{equation}
where $(\xi^\prime(\mathcal{D}_n),\lambda^\prime(\mathcal{D}_n))$ is defined in terms of the 
sample of observations $\mathcal{D}_n$ \citep[Section~3.3.3]{robert:2007}.
Equation (\ref{eq:conpri}) simply says that the conjugate prior is such that
the prior and posterior densities belong to the same parametric family
of densities but with different parameters. In this conjugate setting, it is the parameters
of the posterior density themselves that are ``updated'', based on the observations, relative
to the prior parameters, instead of changing the whole shape of the distribution. To avoid confusion, the parameters
involved in the prior distribution on the model parameter are usually called {\em hyperparameters}.
(They can themselves be associated with prior distributions, then called {\em hyperpriors}.)

The computation of estimators, of confidence regions or of other types of summaries of 
interest on the conjugate posterior distribution often becomes straightforward. 

As a first  illustration, note that a conjugate family of priors for the Poisson model is the collection of gamma 
distributions $\Gamma(a,b)$, since
$$
f_\theta(x) \pi(\theta|a,b) \propto \theta^{a-1+x} e^{-(b+1)\theta} 
$$
leads to the posterior distribution of $\theta$ given $X=x$ being the gamma distribution $\mathcal{G}a(a+x,b+1)$. 
(Note that this includes the exponential distribution $\mathcal{E}xp(2)$ used on the dataset of Figure \ref{fig:cancer}.
The Bayesian estimator of the average survival rate, associated with the quadratic loss, is then given by $\hat{\theta} 
= (1+x_1+x_2+x_3)/(2+N_1+N_2+N_3)$, the posterior mean.)

As a further illustration, consider the case of the normal distribution $\mathscr{N}(\mu,1)$, which is 
indeed another case of an exponential
family, with $\theta=\mu$, $R(x)=x$, and $\Psi(\mu)=\mu^2/2$. The corresponding conjugate prior
for the normal mean $\mu$ is thus normal,
$$
\mathscr{N}\left(\lambda^{-1}\xi ,\lambda^{-1} \right)\,.
$$
This means that, when choosing a conjugate prior in a normal setting, one has
to select both a mean and a variance a priori. (In some sense, this is the
advantage of using a conjugate prior, namely that one has to select only a few parameters
to determine the prior distribution. Conversely, the drawback of conjugate priors is that
the information known a priori on $\mu$ either may be insufficient to determine
both parameters or may be incompatible with the structure imposed by conjugacy.)\index{Conjugacy}
Once $\xi$ and $\lambda$ are selected, the posterior distribution on $\mu$ for a single observation 
$x$ is determined by Bayes' theorem,
\begin{eqnarray*}
\pi(\mu|x) & \propto & \exp(x\mu-\mu^2/2)\,\exp(\xi\mu-\lambda\mu^2/2)\\
           & \propto & \exp\left\{ - (1+\lambda)\left[\mu-(1+\lambda)^{-1}(x+\xi)\right]^2/2\right\}\,,
\end{eqnarray*}
i.e.~a normal distribution with mean $(1+\lambda)^{-1}(x+\xi)$ and variance $(1+\lambda)^{-1}$. 
An alternative representation of the posterior mean is
\begin{equation}\label{eq:nuevo}
\frac{\lambda^{-1}}{1+\lambda^{-1}}\,x + \frac{1}{1+\lambda^{-1}}\,\lambda^{-1}\xi\,,
\end{equation}
that is, a weighted average of the observation $x$ and the prior mean
$\lambda^{-1}\xi$. The smaller $\lambda$ is, the closer the posterior mean is to $x$.
The general case of an iid sample $\mathscr{D}_n=(x_1,\ldots,x_n)$ from the normal distribution $\mathscr{N}(\mu,1)$
is processed in exactly the same manner, since $\bar x_n$ is a sufficient statistic with 
normal distribution $\mathscr{N}(\mu,1/n)$: the $1$'s in \eqref{eq:nuevo} are then replaced with
$n^{-1}$'s.

The general case of an iid sample
$\mathscr{D}_n=(x_1,\ldots,x_n)$  from the normal distribution $\mathscr{N}(\mu,\sigma^2)$ with
an unknown $\theta=(\mu,\sigma^2)$ also allows for a conjugate processing. The normal distribution
does indeed remain an exponential family when both parameters are unknown. It is of the form
$$
(\sigma^2)^{-\lambda_\sigma-3/2}\,\exp\left\{-\left(\lambda_\mu(\mu-\xi)^2+\alpha\right)/2\sigma^2\right\}
$$
since
\begin{eqnarray}\label{eq:conjunor}
\pi((\mu,\sigma^2)|\mathscr{D}_n) & \propto & (\sigma^2)^{-\lambda_\sigma-3/2}\,
	   \exp\left\{-\left(\lambda_\mu (\mu-\xi)^2 + \alpha \right)/2\sigma^2\right\}\nonumber\\
       && \times 
       (\sigma^2)^{-n/2}\,\exp \left\{-\left(n(\mu-\overline{x})^2 + s_x^2 \right)/2\sigma^2\right\} \\
       &\propto& (\sigma^2)^{-\lambda_\sigma(\mathscr{D}_n)}\exp\left\{-\left(\lambda_\mu(\mathscr{D}_n)
           (\mu-\xi(\mathscr{D}_n))^2+\alpha(\mathscr{D}_n)\right)/2\sigma^2\right\}\,,\nonumber
\end{eqnarray}
where $s_x^2 = \sum_{i=1}^n (x_i-\overline{x})^2$.
Therefore, the conjugate prior on $\theta$ is the product of an inverse gamma
distribution on $\sigma^2$, $\mathscr{IG}(\lambda_\sigma,\alpha/2)$, and,
conditionally on $\sigma^2$, a normal distribution on $\mu$, $\mathscr{N} (\xi,\sigma^2/\lambda_\mu)$.

The apparent simplicity of conjugate priors is however not a reason that 
makes them altogether appealing, since there is no further (strong) justification
to their use. One of the difficulties with such families of priors is the influence of the hyperparameter $(\xi,\lambda)$. If the prior
information is not rich enough to justify a specific value of $(\xi,\lambda)$, arbitrarily  fixing $(\xi,\lambda) = (\xi_0,\lambda_0)$
is problematic, since it does not take into account the prior uncertainty on $(\xi_0,\lambda_0)$ itself. To improve on this aspect of conjugate 
priors, a more ameanable solution is to consider a \textit{hierarchical prior}, 
i.e.~to assume that $\gamma=(\xi,\lambda)$ itself is random and to consider a 
probability distribution with density $q$ on $\gamma$, leading to
\begin{eqnarray*}
  \theta|\gamma &\sim& \pi(\theta|\gamma)\\
  \gamma & \sim& q(\gamma)\,,
\end{eqnarray*}
as a joint prior on $(\theta,\gamma)$. The above  is equivalent to considering, as a prior on $\theta$
$$
\pi(\theta) = \int_\Gamma \pi(\theta|\gamma)q(\gamma)\text{d}\gamma\,.
$$
As a general principle,
$q$ may also depend on some further hyperparameters $\eta$. Higher order levels in the hierarchy are thus possible,
even though the influence of the hyper(-hyper-)parameter $\eta$ on the posterior distribution of $\theta$ is
usually smaller than that of $\gamma$. But multiple levels are nonetheless useful in complex populations as those
found in animal breeding \citep{sorensen:gianola:2002}.

Instead of using conjugate priors, even when mixed with hyperpriors, 
one can opt for the so-called {\em noninformative} (or {\em vague\/}) priors \citep{robert:chopin:rousseau:2009}
in order to attenuate the impact on the resulting inference.\index{Prior!noninformative} These priors are 
defined as refinements of the uniform distribution, which 
rigorously does not exist on unbounded spaces. A peculiarity of those vague priors is indeed
that their density usually fails to integrate to one since they have infinite mass, i.e.
$$\int_\Theta \pi(\theta)\text{d}\theta = +\infty,$$
and they are defined instead as positive measures, the first and foremost example being the
Lebesgue measure on $\mathbb{R}^p$. While this sounds like an invalid extension
of the standard probabilistic framework---leading to their denomination of {\em improper
priors}---, it is quite correct to define the corresponding
posterior distributions by \eqref{eq:poste}, provided the integral in the denominator is defined, i.e.
$$
\int \pi(\theta)\ell(\theta|\mathscr{D}_n)\,\hbox{d}\theta =m(\mathscr{D}_n) < \infty\,.
$$
In some cases, this difficulty disappears when the sample size $n$ is large enough.
In others like mixture models (see also Section \ref{sub:test}), the impossibility
of using a particular improper prior may remain whatever the sample size is. It is
thus strongly advised, when using improper priors in new settings, to check that the 
above finiteness condition holds.

The purpose of noninformative priors is to set a prior reference that has very little bearing on the inference
(relative to the information brought by the likelihood function). More detailed accounts are provided in 
\citet[Section~1.5]{robert:2007} about this possibility of using $\sigma$-finite measures
in settings where genuine probability prior distributions are too difficult to come by or too subjective to be accepted by all. 

While a seemingly natural way of constructing noninformative priors would
be to fall back on the uniform (i.e.~flat) prior, this solution has many drawbacks, 
the worst one being that it is not invariant under a change of
parameterisation. To understand this issue, consider the example of a Binomial model: 
the observation $x$ is a $\mathcal B(n,p)$ random variable,
with $p \in (0,1)$ unknown. The uniform prior $\pi(p) = 1$ could then sound like the 
most natural noninformative choice; however, if, instead of the mean parameterisation 
by $p$, one considers the logistic parameterisation $\theta = \log (p/(1-p))$
then the uniform prior on $p$ is transformed into the logistic density
$$
\pi(\theta) = e^\theta / (1 + e^\theta)^2
$$
by the Jacobian transform, which is not uniform. There is therefore a lack of invariance under reparameterisation,
which in its turn implies that the choice of the parameterisation associated with the uniform prior is influencing 
the resulting posterior. This is generaly considered to be a drawback. Flat priors are therefore mostly restricted 
to {\em location models} $x \sim p(x-\theta)$, while {\em scale models}
$$
  x \sim p({x}/{\theta})/\theta
$$
are associated with the log-transform of a flat prior, that is,
$$
   \pi(\theta) = 1/\theta\,.
$$
In a more general setting, the (noninformative) prior favoured by most Ba\-ye\-si\-ans 
is the so-called {\em Jeffreys'} (\citeyear{jeffreys:1939}) {\em prior},
which is related to Fisher's information $I^F(\theta)$ by
$$
\pi^J(\theta) = \left| I^F(\theta) \right|^{1/2}\,,
$$
where $|I|$ denotes the determinant of the matrix $I$.

\begin{figure}
\includegraphics[height=\textwidth,width=5truecm,angle=270]{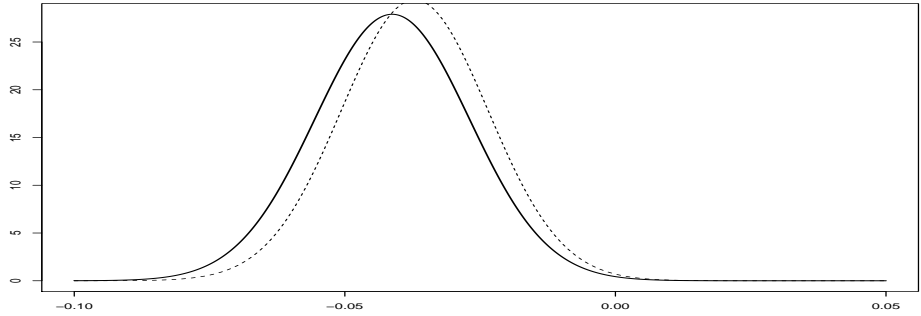}
\caption{Two posterior distributions on a normal mean corresponding to the flat prior (plain)
and a conjugate prior (dotted) for a dataset of 90 observations. {\em (Source: \citealp{marin:robert:2007}.)}}
\label{fig:2nopo}
\end{figure}

Since the mean $\mu$ of a normal model $\mathcal{N}(\mu,1)$ is a location parameter, the standard choice
of noninformative prior is then $\pi(\mu)=1$ (or any other constant).
Given that this flat prior formally corresponds to the choice $\lambda=\mu=0$
in the conjugate prior, it is easy to verify that
this noninformative prior is associated with the posterior distribution $\mathscr{N}(x,1)$.
An interesting consequence of this remark is that the posterior density (as a function of the parameter $\theta$)
is then equal to the likelihood function, which shows that Bayesian analysis subsumes likelihood analysis in this
sense. Therefore, the MAP estimator is also the maximum 
likelihood estimator in that special case. Figure \ref{fig:2nopo} provides the posterior distributions
associated with both the flat prior on $\mu$ and the conjugate 
        $\mathscr{N}(0,0.1\,\hat\sigma^2)$
prior for a crime dataset discussed in \cite{marin:robert:2007}. The difference between both posteriors
is still visible after 90 observations and it illustrates the impact of the choice of 
the hyperparameter $(\xi,\lambda)$ on the resulting inference. 

\subsection{Confidence intervals}\label{sub:confit}

As should now be clear, the Bayesian approach is a complete inferential approach. Therefore, 
it covers among other things confidence evaluation, testing, prediction, model checking, 
and point estimation. Unsurprisingly, the derivation of the confidence intervals (or 
of confidence regions in more general settings)
is based on the posterior distribution $\pi(\theta|\mathscr{D}_n)$. Since the
Bayesian approach processes $\theta$ as a random variable and conditions upon the
observables $\mathscr{D}_n$, a natural definition
of a confidence region on $\theta$ is to determine $C(\mathscr{D}_n)$ such that
\begin{equation}\label{eq:cove}
  \pi(\theta\in C(\mathscr{D}_n) | \mathscr{D}_n) = 1-\alpha
\end{equation}
where $\alpha$ is either a predetermined level such as $0.05$.\footnote{There 
is nothing special about $0.05$ when compared
with, say, $0.87$ or $0.12$. It is just that the famous $5$\%~level is adopted by most as an
acceptable level of error.} or a value derived from the loss function (that may depend on the data).

The important difference from a traditional perspective is that the integration here is done over the
parameter space, rather than over the observation space. The quantity $1-\alpha$
thus corresponds to the probability that a random $\theta$ belongs to this set
$C(\mathscr{D}_n)$, rather than to the probability that the random set contains
the ``true" value of $\theta$. Given this drift in the interpretation of a
confidence set (rather called a {\em credible set} by Bayesians in order to stress this
major difference with the classical confidence set),
the determination of the best\footnote{In the sense of offering a given
confidence coverage for the smallest possible length/volume.} confidence 
set turns out to be easier than in the
classical sense: It simply corresponds to the values of $\theta$ with the
highest posterior values,
$$
C(\mathscr{D}_n)=\left\{\theta;\,\pi(\theta|\mathscr{D}_n) \ge k_\alpha \right\}\,,
$$
where $k_\alpha$ is determined by the coverage constraint \eqref{eq:cove}.
This region is called the {\em highest posterior density} (HPD) region.\index{HPD region}

\begin{figure}
\begin{minipage}[b]{0.5\linewidth}
\includegraphics[width=\textwidth]{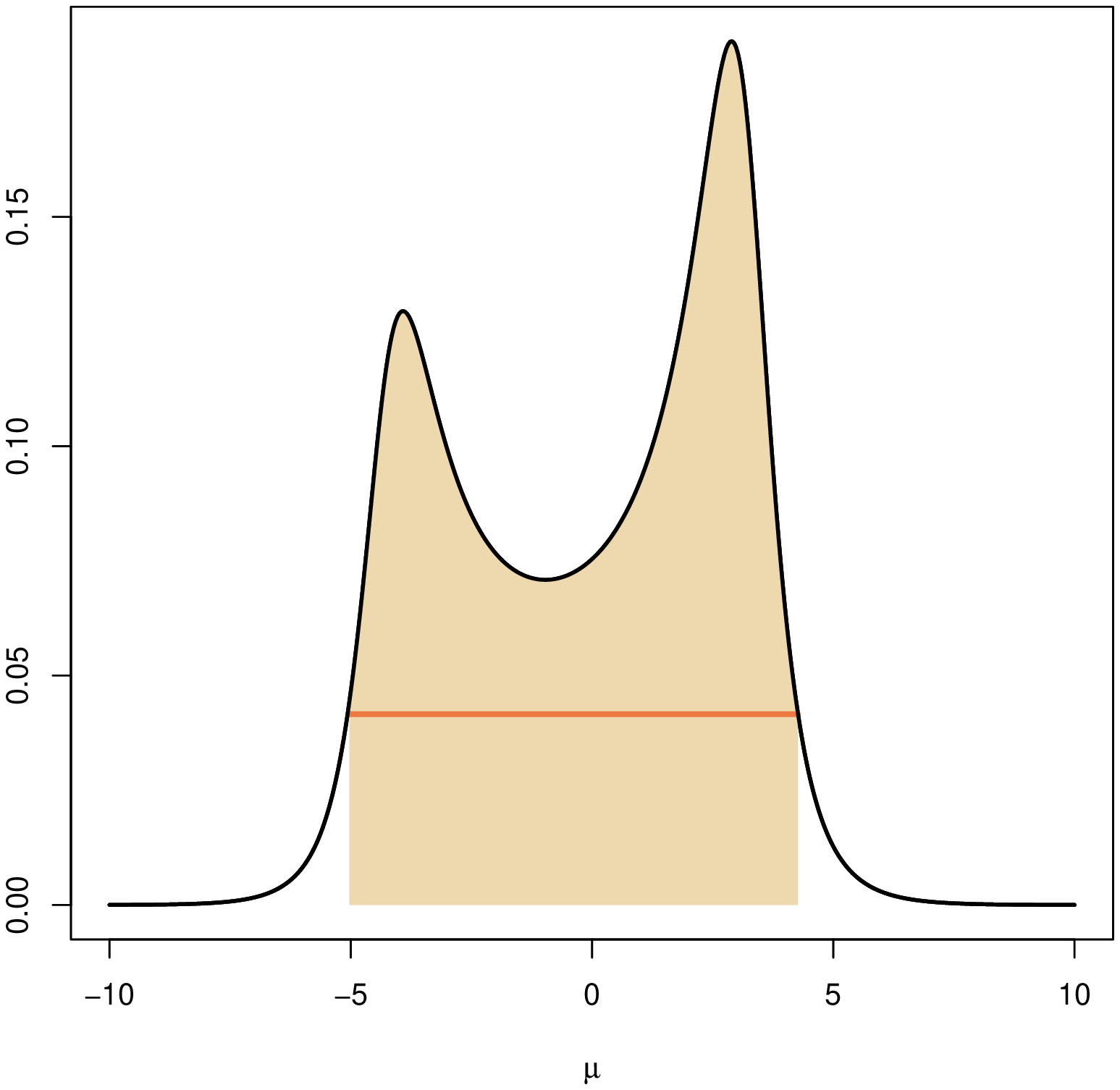}
\end{minipage}
\begin{minipage}[b]{0.5\linewidth}
\includegraphics[width=\textwidth]{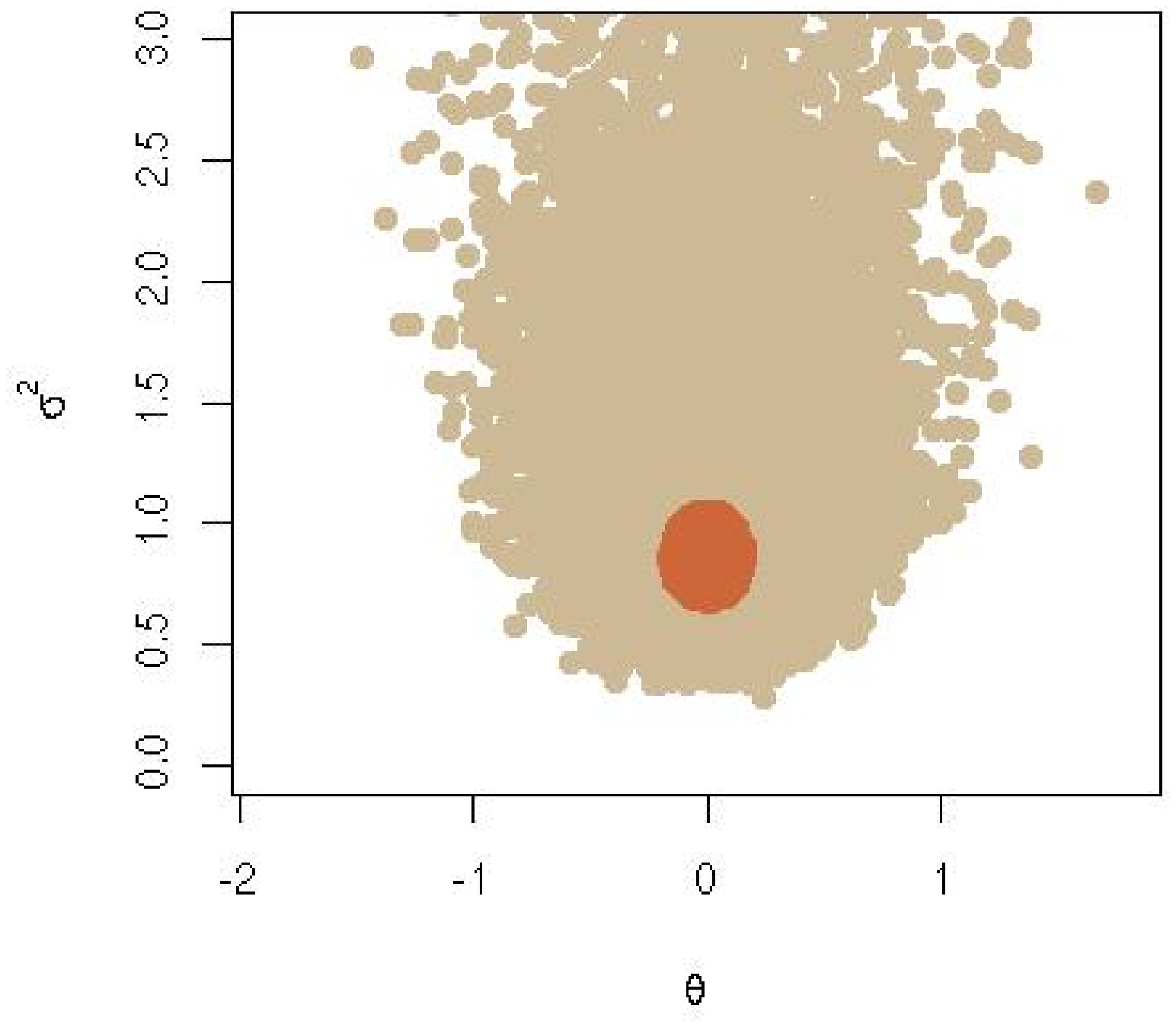}
\end{minipage}
\caption{{\em (left)} Posterior distribution of the location parameter $\mu$ of a Cauchy sample 
for a $\mathscr{N}(0,10)$ prior and corresponding $95\%$ HPD region
({\em Source:} \citealp{marin:robert:2007});
{\em (right)} Representation of a posterior sample of $10^3$ values of 
$(\theta,\sigma^2)$ for the normal model, $x_1,\ldots,x_{10}\sim\mathcal{N}(\theta,\sigma^2)$ with
$\overline x=0$, $s^2=1$ and $n=10$, under Jeffreys' prior,
along with the pointwise approximation to the $10\% $ HPD region {\em (in darker hues)} 
({\em Source:} \citealp{robert:wraith:2009}).}
\label{fig:2cau}
\end{figure}

When the prior distribution is not conjugate, the posterior distribution
is not necessarily so easily-managed.  For instance, if the normal $\mathscr{N}(\mu,1)$ distribution
is replaced with the Cauchy distribution, $\mathscr{C}(\mu,1)$, in
the likelihood
$$
\ell(\mu|\mathscr{D}_n)=\prod_{i=1}^nf_\mu(x_i)={1}\bigg/{\pi^n\prod_{i=1}^n(1+(x_i-\mu)^2)}\,,
$$
there is no conjugate prior available and we can consider a normal prior on $\mu$, say $\mathscr{N}(0,10)$.
The posterior distribution is then proportional to
$$
\tilde\pi(\mu|\mathscr{D}_n)={\exp(-\mu^2/20)}\bigg/{\prod_{i=1}^n(1+(x_i-\mu)^2)}\,.
$$
Solving $\tilde\pi(\mu|\mathscr{D}_n)=k$
is not possible analytically, only numerically, and the derivation of the
bound $k_\alpha$ requires some amount of trial-and-error
in order to obtain the correct coverage. Figure \ref{fig:2cau} gives the
posterior distribution of $\mu$ for the observations $x_1=-4.3$ and $x_2=3.2$.
For a given value of $k$, a trapezoidal approximation can be used to
compute the approximate coverage of the HPD region. For $\alpha=0.95$, a trial-and-error exploration
of a range of values of $k$ then leads to an approximation of $k_\alpha=0.0415$ and the
corresponding HPD region is represented in Figure \ref{fig:2cau} {(\em (left)}.

As illustrated in the above example,
posterior distributions are not necessarily unimodal and thus
the HPD regions may include several disconnected sets.
This may sound counterintuitive from a classical
point of view, but it must be interpreted as indicating indeterminacy,
either in the data or in the prior, about the possible values of $\theta$.
Note also that HPD regions are dependent on the choice of the reference
measure that defines the volume (or surface).\index{Volume}

The analytic derivation of HPD regions is rarely straightforward but let us stress that, due to the fact that
the posterior density is most known up to a normalising constant, those regions can be easily derived from
posterior simulations. For instance,  Figure \ref{fig:2cau} {\em (right)} illustrates this derivation in the 
case of a normal $\mathcal{N}(\theta,\sigma^2)$ model with both parameters unknown and Jeffreys' prior,
when the sufficient statistics are $\overline x=0$ and $s^2=1$, based on $n=10$ observations.

\section{Testing Hypotheses}\label{sub:test}\index{Testing|(}\index{Hypothesis testing}

Deciding about the validity of some restrictions on
the parameter $\theta$ or on the validity of a whole model---like
whether or not the normal distribution is appropriate for the data
at hand---is a major and maybe the most important component of statistical inference.
Because the outcome of the decision process is clearcut, {\em accept} (coded by {\tt 1}) or 
{\em reject\/} (coded by {\tt 0}), the construction and the evaluation of procedures in this 
setup are quite crucial.  While the Bayesian solution is formally very close to a likelihood 
ratio statistic, its numerical values and hence its conclusions often strongly differ from 
the classical solutions.

\subsection{Decisions}\label{sub:0/1}

Without loss of generality, and including the setup of model choice,
we represent null hypotheses as restricted parameter spaces,
namely $\theta \in\Theta_0$. For instance, $\theta>0$ corresponds to
$\Theta_0=\mathbb{R}^+$. The evaluation of testing procedures can be
formalised via the $0-1$ loss that equally penalizes all errors: 
If we\index{Loss function!$0-1$}
consider the test of $H_0:\ \theta \in\Theta_0$ versus $H_1:\ \theta
\not\in\Theta_ 0$, and denote by $d\in\{0,1\}$ the decision made
by the researcher and by $\delta$ the corresponding decision procedure, the loss
$$
L (\theta ,d)  =  \begin{cases} 1-d & \hbox{if}\quad \theta \in \Theta_0\,, \cr
                          d         &  \hbox{otherwise,} \cr \end{cases}
$$
is associated with the Bayes decision (estimator) 
$$
\delta^\pi(x)  =  \begin{cases} 1 & \hbox{if}\quad P^\pi(\theta 
			\in \Theta_0|x)>P^\pi(\theta \not \in \Theta_0|x), \cr
                                0 & \hbox{otherwise.}\cr \end{cases}
$$
This estimator is easily justified on an intuitive basis since it chooses the hypothesis with the 
largest posterior probability. The Bayesian testing procedure is therefore a direct transform of
the posterior probability of the null hypothesis.

\subsection{The Bayes Factor}\label{sub:bafa}

A notion central to Bayesian testing is the {\em Bayes factor}\index{Bayes factor}
$$
B^\pi_{10} = \dfrac{ P^\pi(\theta \in \Theta_1|x) / P^\pi(\theta \in \Theta_0|x) }
{ P^\pi(\theta \in \Theta_1)/ P^\pi(\theta \in \Theta_0) }\,,
$$
which corresponds to the classical odds or likelihood ratio, the difference being that
the parameters are integrated rather than maximized under each model. While it is a
simple one-to-one transform of the posterior probability, it can be used for Bayesian
testing without resorting to a specific loss, evaluating the strength of the evidence in
favour or against $H_0$ by the distance of $\log_{10} (B^\pi_{10})$ from zero \citep{jeffreys:1939}.
This somehow {\em ad-hoc} perspective provides a
reference for hypothesis assessment with no need to define the prior probabilities
of $H_0$ and $H_1$, which is one of the advantages of using the Bayes factor.
In general, the Bayes factor does depend on prior information, but it can be
perceived as a Bayesian likelihood ratio since, if $\pi_0$ and
$\pi_1$ are the prior distributions under $H_0$ and
$H_1$, respectively, $B^\pi_{10}$ can be written as
$$
B^\pi_{10}=\frac{\int_{\Theta_1}f_\theta(x)\pi_1(\theta)\,\hbox{d}\theta}
{\int_{\Theta_0}f_\theta(x)\pi_0(\theta)\,\hbox{d}\theta}=\frac{m_1(x)}{m_0(x)}\,,
$$
thus replacing the likelihoods with the marginals under both hypotheses. Thus,
by integrating out the parameters within each hypothesis, the uncertainty on each 
parameter is taken into account, which induces a natural penalisation for larger models,
as intuited by \cite{jeffreys:1939}. The Bayes factor is 
connected with the Bayesian information criterion (BIC, see \citealp{robert:2007}, Chapter 5), with a penalty term of the form
$d\log n/2$, which explicits the penalisation induced by Bayes factors in regular parametric models. 
In a wide generality, the Bayes factor asymptotically corresponds to a likelihood ratio
with a penalty of the form $d^* \log n^* /2$ where $d^*$ and $n^*$ can be viewed as the {\em effective} dimension of the model and number
of observations, respectively, see \citep{berger:ghosh:2003,chambaz:rousseau:2008}. The Bayes factor therefore offers the major interest
that it does not require to compute a complexity measure (or penalty term)---in other words, to define what is $d^*$ and what is  $n^*$---,
which often is quite complicated and may depend on the true distribution.

\subsection{Point null hypotheses}\label{sub:pointZ}
When the hypothesis to be tested is a point null hypothesis, $H_0:\theta=\theta_0$,
there are difficulties in the construction of the Bayesian procedure, given that, for
an absolutely continuous prior $\pi$,
$$
P^\pi(\theta=\theta_0)=0\,.
$$
Rather logically, point null hypotheses can be criticized as being artificial and impossible
to test ({\em how often can one distinguish $\theta=0$ from $\theta=0.0001$?!\/}), but they
must also be processed, being part of the everyday requirements of statistical analysis
and also a convenient representation of some model choice problems (which we will
discuss later).

Testing point null hypotheses actually requires a modification of the prior
distribution so that, when testing $H_0:\theta\in\Theta_0$ versus $H_1:\theta\in\Theta_1$,
$$
\pi(\Theta_0)>0 \quad\hbox{and}\quad \pi(\Theta_1)>0
$$
hold, whatever the measures of $\Theta_0$ and $\Theta_1$ for the original prior,
which means that the prior must be decomposed as
$$
\pi(\theta) = P^\pi(\theta\in\Theta_0)\times\pi_0(\theta) +
              P^\pi(\theta\in\Theta_1)\times\pi_1(\theta)
$$
with positive weights on both $\Theta_0$ and $\Theta_1$.

Note that this modification makes sense from both informational and
operational points of view. If $H_0:\theta=\theta_0$, the fact that
the hypothesis is tested implies that $\theta=\theta_0$ {\em is} a
possibility and it brings some additional prior information on the
parameter $\theta$. Besides, if $H_0$ is tested {\em and} accepted, this
means that, in most situations, the (reduced) model under $H_0$ will be used
rather than the (full) model considered before. Thus, a prior distribution
under the reduced model must be available for potential later inference. 
(Formaly, the fact that this later inference depends on the selection of $H_0$ 
should also be taken into account.)

In the special case $\Theta_0=\{\theta_0\}$, $\pi_0$ is the Dirac mass at
$\theta_0$, which simply means that $P^{\pi_0}(\theta=\theta_0)=1$, and
we need to introduce a separate prior weight of $H_0$, namely,
$$
\rho = P^\pi (\theta=\theta_0) \quad\hbox{and}\quad
\pi(\theta) = \rho \mathbb{I}_{\theta_0}(\theta)+ (1-\rho) \pi_1(\theta)\,.
$$
Then,
$$
\pi(\Theta_0|x) = \dfrac{f_{\theta_0}(x)\rho }{
                \int f_\theta(x)\pi(\theta)\,\hbox{d}\theta} 
 = \dfrac{f_{\theta_0}(x)\rho }{ f_{\theta_0}(x)\rho + (1-\rho) m_1(x)}.
$$

In the case when $x\sim \mathcal{N}(\mu,\sigma^2)$ and $\mu\sim\mathcal{N}(\xi,\tau^2)$,
consider the test of $H_0:\mu=0$. We can choose $\xi$ equal to $0$ if we
do not have additional prior information. Then the Bayes factor is
the ratio of marginals under both hypotheses, $\mu=0$ and $\mu\ne 0$,
$$
B^\pi_{10} =  {m_1(x)\over f_0(x)} =  \dfrac{\sigma}{ \sqrt{\sigma^2+\tau^2}}\,
        \dfrac{e^{-x^2/2(\sigma^2+\tau^2)} }{ e^{-x^2/2\sigma^2}} 
$$
and
$$
\pi(\mu=0|x) = \left[ 1+{1-\rho\over \rho} \sqrt{{\sigma^2\over
\sigma^2+\tau^2}} \exp\left({\tau^2x^2\over 2\sigma^2(\sigma^2+\tau^2)}\right)
\right]^{-1}
$$
is the posterior probability of $H_0$. Table \ref{tab:nortes} gives an
indication of the values of the posterior probability when the normalized
quantity $x/\sigma$ varies. This posterior probability again depends on
the choice of the prior variance $\tau^2$: The dependence is actually quite
severe, as shown below with the {\em Jeffreys--Lindley paradox}.
\begin{table}
\begin{small}
\begin{tabular}{c c c c c}
\toprule
$ z $ & $ 0$ & $ 0.68 $ & $ 1.28$ & $ 1.96 $ \\
\midrule
$ \pi(\mu=0|z)\ \ \ $ & $\ 0.586$ \ \ & $\ 0.557 \ \ $ & $\ 0.484$ \ \ & $\ 0.351 $ \\
$ \pi(\mu=0|z)\ $ & $\ 0.768$ & $\ 0.729 $ & $\ 0.612$ & $\ 0.366 $ \\
\botrule
\end{tabular}
\caption{Posterior probability of $\mu=0$ for different
values of $z=x/\sigma$, $\rho=1/2$, and for $\tau=\sigma$ (top), 
$\tau^2=10 \sigma^2$ (bottom).\label{tab:nortes}}{{\em Source: \citealp{marin:robert:2007}.}}
\end{small}
\end{table}

\begin{figure}
\centerline{\includegraphics[width=7truecm,angle=270,draft=F]{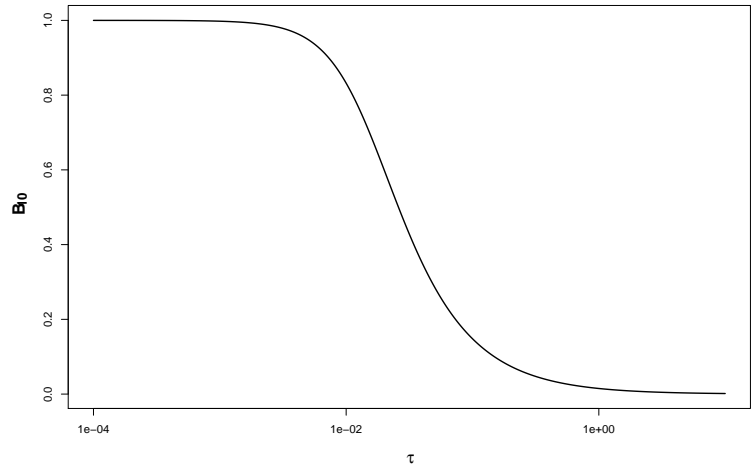}}
\caption{Range of the Bayes factor $B^\pi_{10}$ when $\tau$ 
goes from $10^{-4}$ to $10$. ({\em Note:} The $x$-axis is in logarithmic scale.)
{\em (Source: \citealp{marin:robert:2007}.)}}\label{fig:bfranj}
\end{figure}

\subsection{The Ban on Improper Priors}\label{sub:00}

Unfortunately, this decomposition of the prior distribution into
two subpriors brings a serious difficulty related to improper
priors\index{Prior!improper}, which amounts in practice to banning their use
in testing situations. In fact, when using the representation
$$
\pi(\theta) = P^\pi (\theta\in\Theta_0)\times\pi_0(\theta) +
              P^\pi (\theta\in\Theta_1)\times\pi_1(\theta)\,,
$$
the weights $P^\pi (\theta\in\Theta_0)$ and $P^\pi (\theta\in\Theta_1)$
are meaningful only if $\pi_0$ and $\pi_1$ are normalized probability densities.
Otherwise, they cannot be interpreted as {\em weights}.

In the instance when $x\sim \mathcal{N}(\mu,1)$ and $H_0:\ \mu=0$,
the improper (Jeffreys) prior is $\pi_1(\mu)=1$; if we write
$$
\pi(\mu)  = {1\over 2} \mathbb{I}_{0}(\mu) + {1\over 2} \cdot \mathbb{I}_{\mu\ne 0}\,,
$$
then the posterior probability is
$$
\pi(\mu=0|x) = {e^{-x^2/2}\over e^{-x^2/2}+\int_{-\infty}^{+\infty}
   e^{-(x-\theta)^2/2} \,\hbox{d}\theta} = {1 \over 1+\sqrt{2\pi} e^{x^2/2} }\,.
$$
A first consequence of this choice is that the posterior probability of
$H_0$ is bounded from above by
$$
\pi(\mu=0|x) \le 1/(1+\sqrt{2\pi})=0.285\,.
$$
Table \ref{tab:nortes2} provides the evolution of this probability as $x$
goes away from $0$. An interesting point is that the numerical values somehow
coincide with the $p$-values used in classical testing \citep{Casella:Berger:1990}.

\begin{table}
\begin{small}
\begin{tabular}{c c c c c c}
\toprule
$x $ & $ 0.0$ & $ 1.0 $ & $ 1.65$ & $ 1.96 $ & $ 2.58 $ \cr
\midrule
$ \pi(\mu=0| x)\ $ & $\ 0.285 $ &$\ 0.195$ & $\ 0.089 $ & $\ 0.055$ & $\ 0.014$ \cr
\botrule
\end{tabular}
\end{small}
\caption[Posterior probability\ of $H_0: \mu=0$]{
Posterior probability\ of $H_0: \mu=0$ for the Jeffreys prior $\pi_1(\mu)=1$ under
$H_1$.\label{tab:nortes2}}{{\em Source: \citealp{marin:robert:2007}.}}
\end{table}

If we are instead testing $H_0:\ \theta\le 0$ versus $H_1: \theta>0$,
then the posterior probability is
$$
\pi(\theta\le 0|x) =  {1\over \sqrt{2\pi} } \int_{-\infty}^0
                         e^{-(x-\theta)^2/2} \,\hbox{d}\theta
                   =  \Phi(-x)\,,
$$
and the answer is now {\em exactly} the $p$-value found in classical statistics.

The difficulty in using an improper prior also relates to
what is called the {\em Jeffreys--Lindley paradox}, a phenomenon that shows that
limiting arguments are not valid in testing settings. In contrast with
estimation settings, the noninformative prior no longer corresponds to the
limit of conjugate inferences. In fact, for a conjugate prior, the posterior probability
$$
\pi(\theta=0|x) = \left\{ 1+{1-\rho_0\over \rho_0}
        \sqrt{{\sigma^2\over \sigma^2+\tau^2}} \exp
  \left[{\tau^2x^2\over 2\sigma^2(\sigma^2+\tau^2)}\right] \right\}^{-1}
$$
converges to $1$ when $\tau$ goes to $+\infty$, for {\em every} value of $x$,
as already illustrated by Figure \ref{fig:bfranj}.
This noninformative procedure differs from the noninformative
answer $[1+\sqrt{2\pi}\exp(x^2/2)]^{-1}$ above.

The fundamental issue that bars us from using improper priors on one
or both of the sets $\Theta_0$ and $\Theta_1$ is
a normalizing difficulty: If $g_0$ and $g_1$ are measures (rather than probabilities)
on the subspaces $\Theta_0$ and $\Theta_1$, the choice of the normalizing constants
influences the Bayes factor. Indeed, when $g_i$ is replaced by $c_ig_i$ $(i=0,1)$,
where $c_i$ is an arbitrary constant, the Bayes factor is multiplied by $c_0/c_1$.
Thus, for instance, if the Jeffreys prior is flat and $g_0=c_0$, $g_1=c_1$,
the posterior probability\index{Constant!normalizing}
$$
\pi(\theta\in\Theta_0|x) = {\rho_0c_0\int_{\Theta_0} f_\theta(x)\,\hbox{d}\theta
            \over \rho_0c_0\int_{\Theta_0} f_\theta(x)\,\hbox{d}\theta +
           (1-\rho_0)c_1\int_{\Theta_1} f_\theta(x)\,\hbox{d}\theta }
$$
is completely determined by the choice of $c_0/c_1$. This implies, for instance, that the function
$[1+\sqrt{2\pi}\exp(x^2/2)]^{-1}$ obtained earlier has no validity whatsoever.

Since improper priors are an essential part of the Bayesian
approach, there have been many proposals to overcome this ban. Most use a
device that transforms the prior into a proper probability distribution by
using a portion of the data $\mathscr{D}_n$ and then use the other part of
the data to run the test as in a standard situation. The variety of available solutions
is due to the many possibilities of removing the dependence on the choice of
the portion of the data used in the first step. The resulting procedures are called
{\em pseudo-Bayes factors}.  See Robert (\citeyear[][Chapter 5]{robert:2007}) for more details.

\subsection{The case of nuisance parameters}\label{sub:nuiZ}

In some settings, some parameters are shared by both hypotheses (or by both models)
that are under comparison. Since they have the same meaning in each of both 
models, the above ban can be partly lifted and a common improper prior 
can be used on these parameters, in both models.

\newcommand\by{\mathbf{y}}
\newcommand\bX{\mathbf{X}}
For instance, consider a regression model, represented as
\begin{equation}\label{eq:regZ}
\by|\bX,\beta,\sigma \sim \mathcal{N}(\bX\beta,\sigma^2 I_n)\,,
\end{equation}
where $\bX$ denotes the $(n,p)$ matrix of regressors---upon which the whole analysis is conditioned---, 
$\by$ the vector of the $n$ observations, and
$\beta$ is the vector of the regression coefficients. (This is a matrix representation of the repeated
observation of
$$
y_i = \beta_1 x_{i1}+\ldots+\beta_p x_{ip} + \sigma\epsilon_i\,\quad\epsilon_i\sim\mathcal{N}(0,1)\,,
$$
when $i$ varies from $1$ to $n$.) Variable selection in this setup means removing covariates, that is,
columns of $\bX$, that are not significantly contributing to the expectation of $\by$ given $\bX$. In other
words, this is about testing whether or not a null hypothesis like $H_0:\beta_1=0$ holds. From a Bayesian
perspective, a possible non informative prior distribution on the generic regression model \eqref{eq:regZ}
is the so-called Zellner's (\citeyear{zellner:1986}) $g$-prior,
where the conditional\footnote{The fact that the {\em prior} distribution depends on the matrix of regressors
$\bX$ is not contradictory with the Bayesian paradigm in that the whole analysis is conditional on $\bX$. The
potential randomness of the regressors is not accounted for in this analysis.} 
$\pi(\beta|\sigma)$ prior density corresponds to a normal 
$$
\mathcal{N}(0,n\sigma^2(\bX^\text{T}\bX)^{-1})
$$
distribution on $\beta$, $\mathbf{A}^\text{T}$ denoting the transposed matrix associated with $\mathbf{A}$, 
and where a ``marginal" improper prior on $\sigma^2$, $\pi(\sigma^2) = \sigma^{-2}$, is used to complete
the joint distribution. With this default (or reference) prior modelling, and when considering
the submodel corresponding to the null hypothesis $H_0:\beta_1=0$, with parameters $\beta^{(-1)}$ and $\sigma$, we can
use a similar $g$-prior distribution  
$$
\beta^{(-1)}|\sigma,\bX \sim \mathcal{N}(0,n\sigma^2(\bX_{-1}^\text{T}\bX_{-1})^{-1})\,,
$$
where $\bX_{-1}$ denotes the regression matrix missing the column corresponding to the first regressor, and
$\sigma^2\sim\pi(\sigma^2) = \sigma^{-2}$. Since $\sigma$ is a nuisance parameter in this case, we may use the 
improper prior on $\sigma^2$ as {\em common} to all submodels and thus avoid the indeterminacy in the normalising factor of the
prior when computing the Bayes factor
$$
B_{01} = \dfrac{\int f(\by|\beta_{-1},\sigma,\bX) \pi(\beta^{(-1)}|\sigma,\bX_{-1}) \,\text{d}\beta_{-1}\,
\sigma^{-2}\,\text{d}\sigma}{\int f(\by|\beta,\sigma,\bX) \pi(\beta|\sigma,\bX) \text{d}\beta\,\sigma^{-2}\text{d}\sigma}
$$
Figure \ref{fig:from:Core} reproduces a computer output from \cite{marin:robert:2007} that illustrates how this default prior and the corresponding
Bayes factors can be used in the same spirit as significance levels in a standard regression model, each Bayes factor being associated with 
the test of the nullity of the corresponding regression coefficient. For instance, only the intercept and the coefficients of $X_1,X_2,X_4,X_5$
are significant. This output mimics the standard {\tt lm R} function outcome in order to 
show that the level of information provided by the Bayesian analysis goes beyond the classical output. 
(We stress that all items in the table of Figure \ref{fig:from:Core} are obtained via closed-form formulae.)
Obviously, this reproduction of a frequentist output is not the whole purpose of a Bayesian data anlysis, quite
the opposite: it simply reflects on the ability of a Bayesian analysis to produce automated summaries, just as in
the classical case, but the inferencial abilities of the Bayesian approach are considerably wider. (For instance, 
testing simultaneously the nullity of $\beta_3,\,\beta_6,\ldots,\beta_{10}$ is of identical difficulty, as detailed
in \citealp{marin:robert:2007}, Chapter 3.)

\begin{figure}
\begin{small}
{\sffamily

\begin{tabular}{l l l l}
            &Estimate  &BF        &log10(BF)\\
& & & \\
(Intercept)  &9.2714   &26.334  &1.4205 (***) \\
X1          &-0.0037   &7.0839  &0.8502 (**) \\
X2          &-0.0454   &3.6850  &0.5664 (**) \\
X3          &0.0573    &0.4356  &-0.3609 \\
X4          &-1.0905   &2.8314  & 0.4520 (*) \\
X5          & 0.1953   &2.5157  & 0.4007 (*) \\
X6          &-0.3008   &0.3621  &-0.4412 \\
X7          &-0.2002   &0.3627  &-0.4404 \\
X8          & 0.1526   &0.4589  &-0.3383 \\
X9          &-1.0835   &0.9069  &-0.0424 \\
X10         &-0.3651   &0.4132  &-0.3838 \\
\end{tabular}

\medskip
evidence against H0: (****) decisive, (***) strong,
(**) substantial, (*) poor
}
\end{small}
\caption{{\sf R} output of a Bayesian regression analysis on a processionary caterpillar
dataset with ten covariates analysed in \cite{marin:robert:2007}. The Bayes factor on each row corresponds to
the test of the nullity of the corresponding regression coefficient.}\label{fig:from:Core} 
\end{figure}

\section{Extensions}\label{sec:norex}

The above description of inference is only an introduction and is thus not
representative of the wealth of possible applications resulting from a
Bayesian modelling. We consider below two extensions inspired from \cite{marin:robert:2007}. 

\subsection{Prediction}\label{sub:Nostradamus}\index{Prediction|(}

When considering a sample $\mathscr{D}_n=(x_1,\ldots,x_n)$ from a
given distribution, there can be a
sequential or dynamic structure in the model that implies that
future observations are expected. While more realistic modeling may
involve probabilistic dependence between the $x_i$'s,
we consider here the simpler setup of
predictive distributions in iid settings.

If $x_{n+1}$ is a future observation from the same distribution $f_\theta(\cdot)$ as
the sample $\mathscr{D}_n$,\index{Distribution!predictive}
its {\em predictive distribution} given the current sample is defined as
$$
f^\pi(x_{n+1}|\mathscr{D}_n)  = \int f(x_{n+1}|\theta,\mathscr{D}_n) \pi(\theta|\mathscr{D}_n)\,\hbox{d}\theta
= \int f_\theta(x_{n+1}) \pi(\theta|\mathscr{D}_n)\,\hbox{d}\theta\,.
$$
The motivation for defining this distribution is that the information available on the
pair $(x_{n+1},\theta)$ given the data $\mathscr{D}_n$
is summarized in the joint posterior distribution
$f_\theta(x_{n+1}) \pi(\theta|\mathscr{D}_n)$ and the predictive distribution
above is simply the corresponding marginal on $x_{n+1}$. This is nonetheless coherent with the
Bayesian approach, which then considers $x_{n+1}$ as an extra unknown.

For the normal $\mathscr{N} (\mu,\sigma^2)$ setup, using a conjugate prior on
$(\mu,\sigma^2)$ of the form
$$
(\sigma^2)^{-\lambda_\sigma-3/2}\,\exp-\left\{ \lambda_\mu(\mu-\xi)^2 + \alpha\right\}/2\sigma^2\,,
$$
the corresponding posterior distribution on $(\mu,\sigma^2)$ given $\mathscr{D}_n$ is
$$
\mathscr{N}\left(\frac{\lambda_\mu\xi+n\overline x_n}{\lambda_\mu+n},
\frac{\sigma^2}{\lambda_\mu+n}\right) \times \mathscr{IG} \left(
\lambda_\sigma+n/2,\left[
\alpha+s^2_x+\frac{n\lambda_\mu}{\lambda_\mu+n}(\overline x - \xi)^2\right]/2 \right)\,,
$$
denoted by
$$
\mathscr{N}\left(\xi(\mathscr{D}_n),\sigma^2/\lambda_\mu(\mathscr{D}_n)
\right) \times \mathscr{IG} \left(
\lambda_\sigma(\mathscr{D}_n),\alpha(\mathscr{D}_n)/2 \right)\,,
$$
and the predictive on $x_{n+1}$ is derived as
\begin{align*}
f^\pi(x_{n+1}|\mathscr{D}_n) &\propto \int (\sigma^2)^{-\lambda_\sigma-2-n/2}\,
\exp-(x_{n+1}-\mu)^2/2\sigma^2 \\
&\qquad \times \exp-\left\{ \lambda_\mu(\mathscr{D}_n)(\mu-\xi(\mathscr{D}_n))^2
+ \alpha(\mathscr{D}_n) \right\}/2\sigma^2\,\hbox{d}(\mu,\sigma^2)\\
&\propto \int (\sigma^2)^{-\lambda_\sigma-n/2-3/2}\,
\exp-\left\{ (\lambda_\mu(\mathscr{D}_n)+1)(x_{n+1}-\xi(\mathscr{D}_n))^2\right.\\
&\qquad
/\lambda_\mu(\mathscr{D}_n) +\alpha(\mathscr{D}_n) \big\}/2\sigma^2\,\hbox{d}\sigma^2\\
&\propto \left[ \alpha(\mathscr{D}_n) + \frac{\lambda_\mu(\mathscr{D}_n)+1}{
\lambda_\mu(\mathscr{D}_n)} (x_{n+1}-\xi(\mathscr{D}_n))^2 \right]^{-(2\lambda_\sigma+n+1)/2}\,.
\end{align*}
Therefore, the predictive of $x_{n+1}$ given the sample $\mathscr{D}_n$ is a Student's $t$
distribution with mean $\xi(\mathscr{D}_n)$ and $2\lambda_\sigma+n$ degrees of freedom. In
the special case of the noninformative prior, $\lambda_\mu=\lambda_\sigma=\alpha=0$ and
the predictive is
$$
f^\pi(x_{n+1}|\mathscr{D}_n)\propto
\left[s_x^2+\frac{n}{n+1}(x_{n+1}-\overline x_n)^2\right]^{-(n+1)/2}\,.
$$
This is again a Student's $t$ distribution with mean $\overline x_n)$, scale $s_x/\sqrt{n}$, 
and $n$ degrees of freedom.

\subsection{Outliers}\label{sub:fringe}

Since normal modeling is often an approximation to the ``real
thing," there may be doubts about its adequacy. As already mentioned
above, we will deal later with the problem of checking that the normal
distribution is appropriate for the whole dataset. Here, we consider
the somehow simpler problem of assessing whether or not each point
in the dataset is compatible with normality. There are many
different ways of dealing with this problem. We choose here to take
advantage of the derivation of the predictive distribution above: If
an observation $x_i$ is unlikely under the predictive distribution
based on the {\em other observations}, then we can argue against its
distribution being equal to the distribution of the other
observations.

For each $x_i\in\mathscr{D}_n$, we consider $f^\pi_i(x|\mathscr{D}^{i}_n)$ as 
being the predictive distribution based on $\mathscr{D}^{i}_n=(x_1,\ldots,x_{i-1},
x_{i+1},\ldots,x_n)$. Considering $f^\pi_i(x_i|\mathscr{D}^{i}_n)$ or the
corresponding cdf $F^\pi_i(x_i|\mathscr{D}^{i}_n)$ (in dimension one) gives
an indication of the level of compatibility of the observation with the sample.
To quantify this level, we can, for instance, approximate the distribution of
$F^\pi_i(x_i|\mathscr{D}^{i}_n)$ as uniform over $[0,1]$ since 
$F^\pi_i(\cdot|\mathscr{D}^{i}_n)$ converges to the true
cdf of the model. Simultaneously checking all $F^\pi_i(x_i|\mathscr{D}^{i}_n)$
over $i$ may signal outliers.

The detection of outliers must pay attention to the {\em Bonferroni
fallacy}, which is that extreme values do occur in large enough samples. This means
that, as $n$ increases, we will see smaller and smaller values of
$F^\pi_i(x_i|\mathscr{D}^{i}_n)$ even if the whole sample is from the same
distribution. The significance level must therefore be chosen in accordance with this
observation, for instance using a bound $a$ on $F^\pi_i(x_i|\mathscr{D}^{i}_n)$
such that
$$
1-(1-a)^n = 1-\alpha\,,
$$
where $\alpha$ is the nominal level chosen for outlier detection.

\subsection{Model choice}\label{sec:mocho}
For model choice, i.e.~when several models are under comparison for the same observation
$$
\mathfrak{ M}_i : x \sim f_i(x|\theta_i)\,, \qquad i \in \mathfrak{I}\,,
$$
where $\mathfrak{I}$ can be finite or infinite, the usual Bayesian answer is similar to the Bayesian tests as described above. 
The most coherent perspective (from our viewpoint) is actually to envision the tests of hypotheses as particular cases of model
choices, rather than trying to justify the modification of the prior distribution criticised by \cite{gelman:2008}. This also
incorporates within model choice the alternative solution of model averaging, proposed by \cite{madigan:raftery:1994}, which
strives to keep all possible models when drawing inference.

The idea behind Bayesian model choice is to construct an overall probability on the collection of models 
$\cup_{i \in \mathfrak{I}} \mathfrak{M}_i$ in the following way: the parameter is $\theta = (i, \theta_i)$, i.e. the model index and given the model index equal to $i$, the parameter $\theta_i$ in model $\mathfrak{M}_i$, then the prior measure on the parameter $\theta$ is expressed as
$$
\text{d}\pi(\theta) = \sum_{i\in \mathfrak{I}} p_i \text{d}\pi_i(\theta_i), \quad \sum_{i\in I_i} p_i = 1.
$$
As a consequence, the Bayesian model selection associated with the 0--1 loss function  
and the above prior is the model that maximises the posterior probability
$$
\pi({\mathfrak M}_i|x) = \dfrac{ \displaystyle{p_i \int_{\Theta_i} f_i(x|\theta_i)
         \pi_i(\theta_i) \text{d}\theta_i}}{
\displaystyle{\sum_j p_j \int_{\Theta_j} f_j(x|\theta_j) \pi_j(\theta_j)
      \text{d}\theta_j}  }
$$
across all models.  Contrary to classical pluggin likelihoods, the marginal likelihoods involved in the above ratio 
do compare on the same scale and do not require the models to be nested. As mentioned in Section \ref{sub:nuiZ} integrating 
out the parameters $\theta_i$ in each of the models takes into account their uncertainty thus the marginal likelihoods 
$ \int_{\Theta_i} f_i(x|\theta_i) \pi_i(\theta_i) \text{d}\theta_i$ are naturally penalised likelihoods. In most parametric setups, 
when the number of parameters does not grow to infinity with the number of observations and when those parameters are identifiable.
the Bayesian model selector as defined above is consistent, i.e. with increasing numbers of observations, the probability 
of choosing the right model goes to $1$.

\bibliographystyle{wiley}

\end{document}